\documentclass[aps,prb,floatfix,twocolumn,showpacs,preprintnumbers,superscriptaddress,amsmath,amssymb]{revtex4-1}

\usepackage{graphicx}  
\usepackage{dcolumn}   
\usepackage{bm}        
\usepackage{stmaryrd}  
\usepackage{multirow}  
\usepackage{color}     
\usepackage{amsmath,amssymb}
\usepackage{enumerate}
\usepackage{float}
\usepackage{mathtools}

\newcommand{\der}[3]{\frac{{\rm d}^{#1} #2}{{\rm d} #3^{#1}}}     
\newcommand{\pder}[2]{\frac{\partial #1}{\partial #2}}            
\newcommand{\pderr}[3]{\frac{\partial^{#1} #2}{\partial #3^{#1}}} 
\newcommand{\ppder}[3]{\frac{\partial^2 #1}{\partial #2\, \partial #3}}  
\newcommand{\fder}[2]{\frac{\delta #1}{\delta #2}}                

\def  \ex               {\hat{\bm e}_x}
\def  \ey               {\hat{\bm e}_y}
\def  \ez               {\hat{\bm e}_z}

\def  \eu               {\hat{\bm e}_u}

\def  \gammag           {{\gamma_{\rm g}}}

\def  \blm              {{\bm m}}

\def  \Heff             {{{\bm H}_{\rm eff}}}

\def  \Hext             {{{\bm H}_{\rm app}}}
\def  \happ             {H_{\rm app}}
\def  \Ms               {{M_{\rm s}}}

\def  \bs               {{\bm s}}

\def  \tor              {{\bm \tau}}

\def  \bOm              {{\bm \Omega}}
\def  \bj               {{\bm j}}
\def  \bnab             {{\bm \nabla}}
\def  \Ham              {{\cal H}}

\def  \bs               {{\hat{\bm s}}}
\def  \mus              {\mu_{\rm S}}
\def  \Dp               {{D_{\perp}}}
\def  \Du               {{D_u}}
\def  \wex              {{w_{\rm ex}}}
\def  \wp               {{w_{\perp}}}
\def  \wu               {{w_u}}
\def  \wz               {{w_{\rm Z}}}
\def  \Kp               {K_{\perp}}
\def  \Ku               {K_u}

\def  \omp              {\Omega_\phi}
\def  \omt              {\Omega_\theta}

\def \muB               {{\mu}_{\rm B}}   
\def \ex                {\hat{\bm e}_x}
\def \ey                {\hat{\bm e}_y}
\def \ez                {\hat{\bm e}_z}
\def \et                {\hat{\bm e}_\theta}
\def \ep                {\hat{\bm e}_\phi}

\begin{document}

\preprint{APS/123-QED}

\title{Static properties and current-induced dynamics of pinned $90$ degree magnetic domain walls under applied fields: an analytic approach}

\author{Pavel Bal\'a\v{z}} \email{balaz@karlov.mff.cuni.cz}
\affiliation{Charles University, Faculty of Mathematics and Physics, Department of Condensed Matter Physics, Ke Karlovu 5, CZ 121 16 Prague, Czech Republic}
\affiliation{IT4Innovations Center, VSB Technical University of Ostrava, 17. listopadu 15, CZ 708 33 Ostrava-Poruba, Czech Republic}
\author{Sampo J. H\"am\"al\"ainen}
\affiliation{NanoSpin, Department of Applied Physics, Aalto University School of Science, P.O. Box 15100, FI-00076 Aalto, Finland}
\author{Sebastiaan van Dijken}
\affiliation{NanoSpin, Department of Applied Physics, Aalto University School of Science, P.O. Box 15100, FI-00076 Aalto, Finland}

\date{\today}

\begin{abstract}
Magnetic domain walls are pinned strongly by abrupt changes in magnetic anisotropy. When driven into oscillation by a spin-polarized current, 
locally pinned domain walls can be exploited as tunable sources of short-wavelength spin waves. Here, we develop an analytical 
framework and discrete Heisenberg model to describe the static and dynamic properties of pinned domain walls with a head-to-tail magnetic structure. 
We focus on magnetic domain walls that are pinned by 90$^\circ$ rotations of uniaxial magnetic anisotropy. 
Our model captures the domain wall response to a spin-transfer torque that is exerted by an electric current. 
Model predictions of the domain wall resonance frequency and its evolution with magnetic anisotropy strength and external magnetic field are compared to micromagnetic simulations. 
\end{abstract}


\maketitle

\section{Introduction}
\label{Sec:Intro}

Magnetic domain walls (DWs) are of great interest for spintronics.~\cite{HOF-15} 
The motion of DWs in magnetic nanowires has attracted particular attention because 
of potential applications in data storage and logic devices.~\cite{Allwood:Science_2005,Parkin:Science_2008,Xu-08} 
Magnetic DWs can be driven by a magnetic field, an electric current,~\cite{YAM-04,Li:PRL_2004,Tatara:SciRep_2008,EMO-13,RYU-13} 
propagating spin waves,~\cite{HAN-09,Yan:PRL_2011,HIN-11} or an electric field.~\cite{Lahtinen:SciRep_2012,FRA-15} 
On the other hand, strong DW pinning at specific locations of a ferromagnetic film offers attractive prospects for magnonics,~\cite{KRA-14} 
where they can be used as spin-wave nanochannels~\cite{GAR-15,WAG-16,TRU-16} 
or monochromatic spin-wave sources.~\cite{VanDeWiele:SciRep_2016,Voto2017:SciRep}
In unpatterned films, DW pinning requires a lateral modulation of magnetic anisotropy. 
Here, the anisotropy boundaries pin the magnetic DWs and an external magnetic field tailors 
their spin structure instead of moving them. Deterministic switching between wide 
and narrow magnetic DWs by a magnetic field has been demonstrated~\cite{Franke:PRB_2012} 
and the energetics of different DW types can drastically alter the magnetization reversal process.~\cite{YOU-13,CAS-15} 
For spin-wave emission, a pinned magnetic DW needs to be driven into oscillation by high-frequency actuation. 
Spin-transfer torques from an ac spin-polarized current can be used to achieve this.~\cite{VanDeWiele:SciRep_2016} 
Magnetic anisotropy boundaries themselves can also act as local spin-wave sources in a microwave magnetic field.~\cite{HAM-17} 
Thus, even if all magnetic DWs are erased by an external bias field, spin waves are still emitted from anisotropy boundaries. 
In this case, dissimilar magnetization precessions in neighboring domains trigger the excitation of spin waves.   

Regular modulations of magnetic anisotropy can be induced by magnetoelectric coupling 
between a ferromagnetic film and a ferroelectric layer. 
In some material systems, the ferroelectric domain pattern is completely transfered to the ferromagnet. 
Full ferroelectric-ferromagnetic domain correlations have been demonstrated in bilayers where 
the ferromagnetic film is exchange-coupled to the canted magnetization of a single-phase 
multiferroic film~\cite{CHU-08,Lebeugle:PRL_2009,Heron:PRL_2011,YOU-13} 
or strain-coupled to the ferroelastic domains of a 
ferroelectric crystal.~\cite{Lahtinen:AdvMat_2011,Chopdekar:PRB_2012,STR-13,Franke:PRL_2014} 
In both material systems, a local uniaxial magnetic anisotropy is induced in the ferromagnetic film. 
The in-plane axis of uniaxial magnetic anisotropy rotates from one domain to the other. 
Since ferroelectric domain boundaries in multiferroic bilayers are only a few nanometer wide, 
the magnetic anisotropy boundaries are nearly abrupt. 
Magnetic domains walls are pinned strongly by such sharp rotations of magnetic anisotropy. 
Besides multiferroic heterostructures, modulations of uniaxial magnetic anisotropy can 
also be realized by local ion irradiation~\cite{TRU-16,TRU-14} 
and thermally-assisted scanning probe lithography.~\cite{ALB-16}
In most of the cited examples, the uniaxial magnetic anisotropy axis rotates by 90$^\circ$. 
In thin ferromagnetic films and zero magnetic field, the anisotropy boundaries thus pin 90$^\circ$ magnetic DWs of the N\'{e}el type.      

In this paper, we provide a theoretical description of a magnetic DW that is pinned by a 90$^\circ$ uniaxial magnetic anisotropy boundary. 
To describe the static and dynamic properties of the DW, we use a 1D model with continuous spatial variables. 
The model allows us to accurately calculate the static deformation of the DW profile in a perpendicular magnetic field. 
Dynamic excitations of the DW are modeled by the inclusion of a spin-transfer torque from an electric current. 
Application of a spin-polarized current moves the DW center away from the anisotropy boundary and tilts the DW magnetization out of the film plane. 
Next, we describe the dynamics of a pinned magnetic DW by using the center and tilting angle of the DW as collective coordinates. 
We derive an expression for the DW resonance frequency and calculate how it varies as a function 
of magnetic anisotropy strength and applied magnetic field. 
For consistency, we compare our analytical results with numerical simulations based on a 1D Heisenberg model and 
the Landau-Lifshitz-Gilbert (LLG) equation as well as with micromagnetic simulations.

The paper is organized as follows. In Sec.~\ref{Sec:Model} we introduce the DW models. 
In Sec.~\ref{Sec:DomainWall} we provide results for the DW profile in zero and non-zero magnetic field. 
Sec.~\ref{Sec:Dynam} studies the effect of an applied electric current. 
First, we develop a model for current-induced DW oscillations in zero magnetic field. 
Then, a model describing the simultaneous action of a magnetic bias field and a spin-polarized current is presented. 
An expression for the DW resonance frequency is derived and numerically studied. Finally, we discuss our results in Sec.~\ref{Sec:Conclusions}.

\section{Model}
\label{Sec:Model}

\begin{figure}[htp!]
  \includegraphics[width=.9\columnwidth]{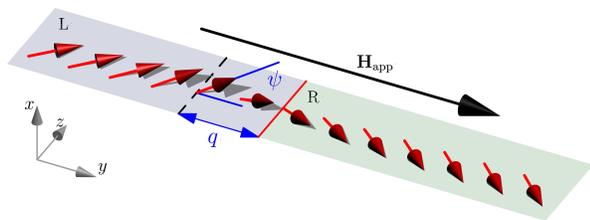}
  \caption{Collective coordinates of an head-to-tail 90$^\circ$ magnetic DW.
  The red arrows point in the direction of local magnetization.
  The left (L) and right (R) parts of the magnetic layer differ in the direction of uniaxial magnetic anisotropy axis.
  The red solid line indicates the abrupt anisotropy boundary and the black dashed line marks the DW center. 
  The displacement of the DW center from the anisotropy boundary is given by coordinate $q$ 
  while the DW tilting angle from the film plane is given by $\psi$. 
  The black arrow indicates the direction of applied magnetic field ($\happ$).}
\label{Fig:collective}
\end{figure}
In our models, we orient the ferromagnetic film in the $y-z$ plane (see Fig.~\ref{Fig:collective}). 
The DW magnetization changes along the $y$-axis and we assume translation symmetry along the $z$-axis. 
The magnetic anisotropy boundary is located at $y = 0$ and the angle between 
the uniaxial anisotropy axis in the left (L) and right (R) domains is set to 90$^\circ$. 
The unit vector along the anisotropy axis in the domains is expressed as $\eu = (0, \sin\xi_u, \cos\xi_u)$, with $\xi_u$ differing in the domains
\begin{equation}
  \xi_u = 
  \begin{dcases}
   \pi/4\,, & \text{in the left domain ($y<0$) }\,, \\
    3\pi/4\,, & \text{in the right domain ($y>0$)}\,.
  \end{dcases}
\label{Eq:xi_u}
\end{equation}

In this section, we introduce two models that describe the static and dynamic properties of a pinned magnetic DW. 
For analytical calculations, we exploit a continuous 1D model in which the magnetization direction varies smoothly across the DW. 
Numerical simulations are performed using a discrete 1D Heisenberg model. 
Here, we consider a finite chain of magnetic moments along the $y$-axis. Relations between the two models are explained.

\subsection{Continuous model}

In the continuous limit, the volume energy density can be written as the sum of exchange energy density ($\wex$), 
shape anisotropy density ($\wp$), Zeeman energy density ($\wz$), and uniaxial anisotropy energy density ($\wu$) 
\begin{equation}
  w = \wex + \wp + \wz + \wu\,.
\end{equation}
Generally, the terms take the forms
\begin{subequations}
  \begin{align}
    \wex &= A\, \left[(\bnab m_x)^2 + (\bnab m_y)^2 + (\bnab m_z)^2 \right]\,, \\
    \wp &= \Kp\, \left(\blm \cdot \ex \right)^2\,, \\
    \wz &= -\mu_0\, \Ms\, \blm \cdot \Hext\,, \\
    \wu &= -\Ku\, \left( \blm \cdot \hat{\bm e}_u \right)^2\,,
  \end{align}
\label{Eqs:w_def}
\end{subequations}
where $\blm = (m_x, m_y, m_z)$ is a unit vector along the magnetization direction $\blm = {\bm M} / \Ms$, 
$\Ms$ is the saturation magnetization, $A$ is the exchange stiffness parameter, 
$\Kp$ is the perpendicular anisotropy, and $\Ku$ is the uniaxial in-plane anisotropy. 
In our calculations, we always assume $\Kp$ and $\Ku$ to be positive.

\subsection{Heisenberg model}

To complement our analytical results, we perform simulations using a discrete Heisenberg model. 
In this model, the continuously varying parameter 
${\bm m}(y)$ is replaced by $\bs_n = {\bm m}(y_n)$, with $y_n$ 
indicating the position of $n$-th spin of a 1D chain. 
The positional variable can be expressed as 
$y_n = (n-1)\, a$, where $n \in \{1, 2, \dots, N\}$ and $a$ is the distance between two adjacent magnetic moments. 
The Heisenberg Hamiltonian of a 1D chain of magnetic moments is given by~\cite{Wieser:PRB_2010}
\begin{equation}
  \begin{split}
  \Ham = &-J \sum_n \bs_n \cdot \bs_{n+1} - \mu_0 \mus\, \Hext \cdot \sum_n \bs_n + \\
         &\Dp \sum_n \left( \bs_n \cdot \ex \right)^2 -
          \Du \sum_n \left( \bs_n \cdot \eu \right)^2\,,
  \end{split}
\label{Eq:Hamiltonian}
\end{equation}
where, $J$ is the exchange coupling parameter, $\Hext$ is the applied magnetic field, $\Dp > 0$ 
is the perpendicular magnetic anisotropy, and $\Du > 0$ is the uniaxial anisotropy in the film plane. 
The parameters in Eq.~\ref{Eq:Hamiltonian} 
are related to those of the continuous model: 
$J = 2a\, A$, $\Dp = a^3\, \Kp$, and $\Du = a^3\, \Ku$, where $a$ is the cell size. 
Moreover, if we define $\mus = \muB S$, where $\muB$ is Bohr magneton and $S$ is the spin per unit cell, 
then the saturated magnetization of one cell is given by $\Ms = \mus / a^3$.

\subsection{Spin dynamics}

We will now describe the dynamics of magnetization that is generated by an effective torque. 
The torques that we consider are caused by an effective magnetic field or a spin-polarized current. 
The time variation of the magnetization vector is described by the Landau-Lifshitz-Gilbert (LLG) equation. 
In the continuous limit, it reads
\begin{equation}
    \der{}{\blm}{t} - \alpha\, \blm \times \der{}{\blm}{t} = \bOm\,, 
\label{Eq:LLG}
\end{equation}
where $t$ is time, $\alpha$ is the Gilbert damping parameter, and $\bOm$ is the total torque acting on $\blm$. $\bOm$ can be written as
\begin{equation}
  \bOm = -\mu_0\, \gamma\, \blm \times \Heff + \tor\,,
\label{Eq:Gamma}
\end{equation}
where $\gamma = |\gammag| > 0$ is the gyromagnetic ratio. 
$\bOm$ consists of two terms, one describing the torque that is induced by the effective magnetic field ($\Heff$) 
and another representing the current-induced spin-transfer torque ($\tor$). 
The effective magnetic field is a functional derivative of the volume energy density ($w$)
\begin{equation}
  \Heff = \frac{1}{\mu_0\,\Ms} \fder{w}{\blm}\,,
\label{Eq:Heff}
\end{equation}
where $\mu_0$ is the vacuum permeability. The spin-transfer torque acting on the magnetization is given by\cite{Li:PRL_2004,Li:PRB_2004}
\begin{equation}
  \tor = -u\, \left[ (\bj \cdot \bnab) - \beta\, \blm \times (\bj \cdot \bnab) \right] \blm\,,
\label{Eq:torque}
\end{equation}
where $\bj$ is a unit vector along the current direction and $\beta$ is the spin-torque nonadiabaticity. The parameter $u$ is given by
\begin{equation}
  u = \frac{\muB I P}{e \Ms}\,,
\label{Eq:u}
\end{equation}
where $I$ is the charge current density, $P$ is the spin polarization of the current,
and $e$ is the electron charge.
If we consider a current along the $y$-axis, we can write
\begin{equation}
  \tor = -u\, \left( \pder{}{y} - \beta\, \blm \times \pder{}{y} \right) \blm.
\label{Eq:torque1}
\end{equation}

\subsubsection{Spherical coordinates}

\begin{figure}
  \centering
  \includegraphics[width=.9\columnwidth]{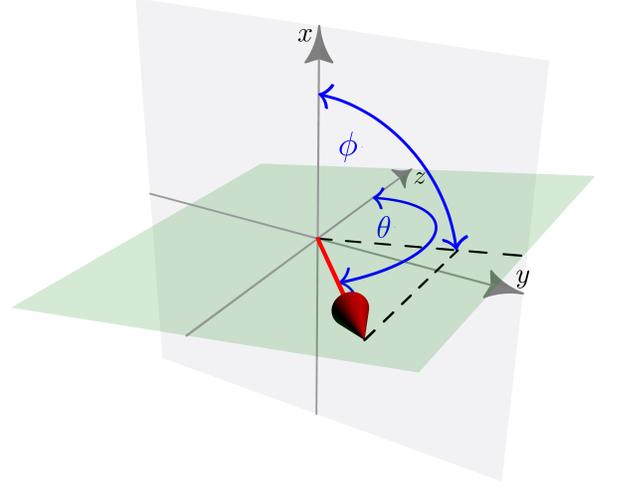}
  \caption{Spherical coordinate system used in the analytical model.}
\label{Fig:spherical}
\end{figure}
For the sake of simplicity, we now express the LLG equation in local spherical coordinates $(\theta, \phi)$, 
as schematically shown in Fig.~\ref{Fig:spherical}. In this coordinate system, the magnetization vector can be written as 
${\bm m} = (\cos\phi \sin\theta, \sin\phi \sin\theta, \cos\theta)$.
Moreover, we define two local base vectors perpendicular to $\blm$.
\begin{subequations}
\label{Eq:loc_base_vectors}
  \begin{align}
    \ep &= \left( \ez \times \blm \right) / \sin\theta\,, \\
    \et &= \ep \times \blm\,,
  \end{align}
\end{subequations}
where $\ez = (0, 0, 1)$. Consequently, the LLG equation (Eq.~\ref{Eq:LLG}) takes the form~\cite{Thiaville:EPL2005,Thomas:Nat2006}
\begin{subequations}
  \begin{align}
    \der{}{\theta}{t} = &-\frac{\gamma}{\Ms} \frac{1}{\sin\theta} \fder{w}{\phi} - \alpha\, \sin\theta\, \der{}{\phi}{t} \notag \\
                        &- u\, \left( \pder{\theta}{y} + \beta\, \sin\theta\, \pder{\phi}{y} \right)\,, \\
    \sin\theta\, \der{}{\phi}{t} = &\quad \frac{\gamma}{\Ms} \fder{w}{\theta} + \alpha\, \der{}{\theta}{t} \notag \\
                                   &- u\, \left( \sin\theta\, \pder{\phi}{y} - \beta\, \pder{\theta}{y} \right)\,.
  \end{align}
\label{Eqs:LLG_spher}
\end{subequations}
The overall torque acting on ${\bm m}$ 
can be split as 
\begin{equation}
  \bOm = \Omega_\theta\, \et + \Omega_\phi\, \ep\,,
\end{equation}
where $\Omega_\theta = \bOm \cdot \et$ and $\Omega_\phi = \bOm \cdot \ep$.

Finally, the different energy density terms can be expressed as
\begin{subequations}
  \begin{align}
    \wex &= A\, \left[ \left( \pder{\theta}{y} \right)^2 + \left( \pder{\phi}{y} \right)^2 \sin^2\theta \right]\,, \\
    \wp &= \Kp\, \cos^2\!\phi\, \sin^2\!\theta\,, \\
    \wz &= -\mu_0\, \Ms\, \happ\, \sin\!\phi\, \sin\!\theta\,, \\
    \wu &= -\frac{\Ku}{\sqrt{2}}\, \left(\sin\phi \sin\theta \pm \cos\theta \right)^2.
  \end{align}
\label{Eqs:w_def_spher}
\end{subequations}
Here, we took into account that the magnetization is solely changing along the $y$ 
direction and the magnetic bias field is oriented along $y$ as well ($\Hext = \happ\, \ey$). 
In the expression for $\wu$, we included the abrupt 90$^\circ$ rotation of uniaxial magnetic anisotropy. 
The upper sign relates to the left domain ($y < 0$) and the lower sign applies to the right domain ($y > 0$).

\subsubsection{Heisenberg model}

In the discrete Heisenberg model, we replace $\blm$ by $\bs_n$ in the LLG equation. 
The effective magnetic field encountered by spin $\bs_n$ is given by
$\Heff_n = -(\mus \mu_0)^{-1} (\delta\Ham/\delta\bs_n)$. 
For the discrete variable we use 
$\partial\bs / \partial y = (\bs_{n+1} - \bs_{n-1}) / (2\,a)$ at the $n$-th site of the 1D chain. 
This gives a discretized expression for the current-induced spin-transfer torque
\begin{equation}
  \tor_n = -\frac{u}{a} \left[ \Delta\bs_n - \beta\, \bs_n \times \Delta \bs_n \right]\,,
\label{Eq:stt_disc}
\end{equation}
where $\Delta \bs_n = (\bs_{n+1} - \bs_{n-1}) / 2$.

\section{90$^\circ$ domain wall}
\label{Sec:DomainWall}

\subsection{Equilibrium DW}

We will now inspect the DW profile in equilibrium, i.e., 
when no external magnetic field and no electric current are applied. 
If we assume that the magnetization rotates in the film plane ($\phi = \pi/2$), $\omt = 0$ in both domains and
\begin{equation}
  \omp = \frac{\gamma}{\Ms} \left\{ -2\, A \pderr{2}{\theta(y)}{y} + K_u \sin\left[2 \left(\theta(y) - \xi_u\right)\right] \right\}\,.
\label{Eq:omp}
\end{equation}
In equilibrium, $\omt = \omp = 0$, which gives~\cite{Tatara:SciRep_2008}
\begin{equation}
  \pderr{2}{\theta'}{y} = \frac{1}{\lambda^2} \sin\theta' \cos\theta'\,.
\label{Eq:dif_eq}
\end{equation}
Here, we defined $\theta'(y) = \theta(y) - \xi_u$ and
\begin{equation}
  \lambda = \sqrt{\frac{A}{K_u}}\,.
\label{Eq:lambda}
\end{equation}
For a head-to-tail 90$^\circ$ DW one needs to impose the boundary conditions $\theta' \to 0$ for $y \to \pm\infty$. 
Moreover, $y=0$ and $\theta = \pi/2$ at the anisotropy boundary. 
Using these conditions, we obtain a static solution for the DW profile
\begin{equation}
  \theta(y) =
    \begin{dcases}
      \frac{\pi}{4} + 2\, \arctan\left[ \left( \sqrt{2} - 1 \right) \exp\left(y/\lambda \right)\right]\,, \\
      \qquad \text{if } y < 0\,, \\
      \frac{3 \pi}{4} - 2\, \arctan\left[ \left( \sqrt{2} - 1 \right) \exp\left(-y/\lambda \right)\right]\,, \\
      \qquad \text{if } y > 0\,.
    \end{dcases}
\label{Eq:prof_eq}
\end{equation}

\begin{figure}[!tp]
 \centering
 \includegraphics[width=.9\columnwidth]{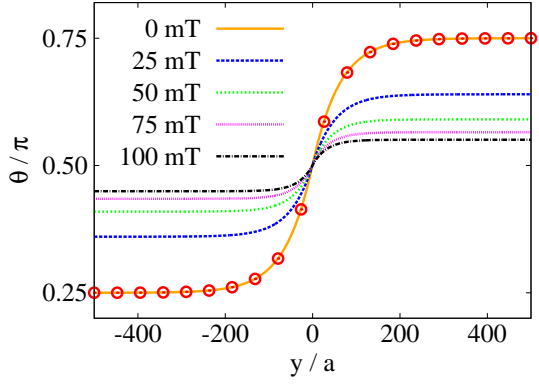}
 \caption{Domain wall profiles obtained from Heisenberg model simulations for different values of applied magnetic field $\mu_0 \happ$. 
          In the simulations, 
          $N = 2000$, $a = 0.5\, {\rm nm}$, $\Ms = 1.5 \times 10^6\, {\rm A/m}$, $A = 2.1 \times 10^{-11}\, {\rm J/m}$, $K_u = 2.5 \times 10^4\, {\rm J/m}^3$, and $\Dp = 0.1 \Du$. 
          The open circles indicate zero-field solutions of the analytic model (Eq.~\ref{Eq:prof_eq}).}
 \label{Fig:DW_prof}
\end{figure}
This expression is exact when dipolar interactions are negligible. 
For a head-to-tail 90$^\circ$ DW this is an accurate approximation because its profile is determined by 
the competing strengths of exchange coupling and uniaxial magnetic anisotropy.~\cite{Franke:PRL_2014} 
Figure~\ref{Fig:DW_prof} demonstrates that the analytical solution agrees well with Heisenberg model 
simulations for zero magnetic field (see solid orange curve and open circles), 
which also ignores dipolar interactions.

When a magnetic field is applied along an unpinned 180$^\circ$ magnetic DW, it moves to minimize Zeeman energy. 
On the other hand, when the field is oriented perpendicular to the same DW, its internal spin structure and, 
thereby, the dynamic properties change.~\cite{Sobolev1994:JAP,Sobolev1995:JMMM} 
Next, we will analyze how the application of a magnetic field normal to the DW plane alters the profile of a pinned 90$^\circ$ DW.

\subsection{Effect of magnetic field}
\label{SSec:mag_field}

When an in-plane magnetic field is applied perpendicular to the head-to-tail 90$^\circ$ DW, i.e., along the $y$-axis, 
the Zeeman energy is the same in both domains. Therefore, the DW will not leave its equilibrium position on top of the anisotropy boundary. 
Instead, the magnetization vectors in both domains gradually rotate towards each other in a magnetic field. 
This coherent reduction of the DW angle depends on the strength of uniaxial magnetic anisotropy.

The torque that acts on the magnetization in an external magnetic field $\happ$ is given by
\begin{equation}
  \begin{split}
    \omp = \frac{\gamma}{\Ms} \biggl\{
	    &-2 A\, \pderr{2}{\theta}{y} - K_u\, \sin\left[2 (\theta - \xi_u)\right] \\
	    &+\happ\, \mu_0 \Ms\, \cos\theta\; \biggr\}\,.
  \end{split}
\label{Eq:omp_happ}
\end{equation}
For $\happ > 0$, the magnetization angle $\theta$ in the left domain increases by angle $\zeta$, $\theta_{\rm L} = \pi/4 + \zeta$, 
while in the right domain $\theta_{\rm R} = 3\pi/4 - \zeta$. 
Consequently, the magnetization rotation between neighboring domains ($\Delta$) is reduced by $2\zeta$; $\Delta = \pi/2 - 2\zeta$.
Figure~\ref{Fig:DW_prof} shows how the DW profile evolves as a function of applied magnetic field.

Deep inside the domains, where $\partial\theta/\partial{y} = 0$, Eq.~\ref{Eq:omp_happ} can be used to derive an expression for $\zeta$
\begin{equation}
  \frac{K_u}{\mu_0\, \Ms} \sin\left(2\, \zeta\right) = \happ \cos\left( \zeta + \frac{\pi}{4} \right)\,.
\label{Eq:zeta}
\end{equation}
This equation can be solved numerically for any arbitrary value of $\happ$. 
Once the angle $\zeta$ is known, one can use the following ansatz for the DW profile in an applied magnetic field
\begin{equation}
  \theta_{\zeta}(y) =
    \begin{dcases}
      \frac{\pi}{4} + \zeta + 2\, \arctan\left[ C\, \exp\left(y/\lambda' \right)\right]\,, \\
      \qquad \text{if } y < 0\,, \\
      \frac{3 \pi}{4} - \zeta - 2\, \arctan\left[ C\, \exp\left(-y/\lambda' \right)\right]\,, \\
      \qquad \text{if } y > 0\,,
    \end{dcases}
\label{Eq:prof_zeta}
\end{equation}
where $C$ can be extracted from the boundary condition at $y=0$
\begin{equation}
  C = \tan\left( \frac{\pi}{8} - \frac{\zeta}{2} \right)\,.
\label{Eq:A_zeta}
\end{equation}
Moreover, $\lambda'$ in Eq.~\ref{Eq:prof_zeta} is the DW width, 
which differs from the zero-field DW width, $\lambda$, as defined by Eq.~\ref{Eq:lambda}.

\begin{figure}[tp]
 \centering
 \includegraphics[width=.9\columnwidth]{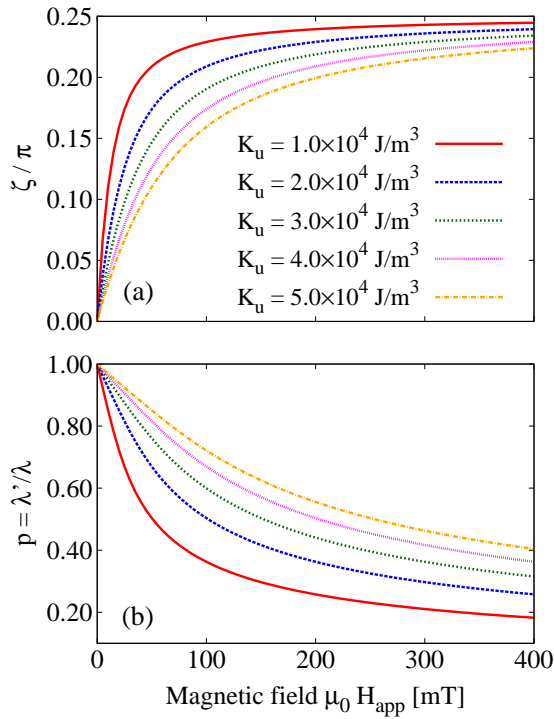}
 \caption{(a) Angle $\zeta$ and (b) $p = \lambda'/\lambda$ 
          as a function of applied magnetic field, $\happ$, for various values of $K_u$. 
          The others parameters in the calculations
          are the same as in Fig.~\ref{Fig:DW_prof}.}
 \label{Fig:dw_happ}
\end{figure}
Figure~\ref{Fig:dw_happ} shows the parameter $\zeta$ and ratio $p = \lambda'/\lambda$ 
as a function of magnetic field for different values of $K_u$. 
While the values of $\zeta$ are directly obtained from Eq.~\ref{Eq:zeta}, 
the dependence of $\lambda'$ follows from Heisenberg model simulations. 
Here, the LLG equation is used to simulate the relaxation of discrete magnetic moments in a magnetic field. 
Once the static state is reached, the parameters $\zeta$ and $\lambda'$ are extracted 
by fitting the spatial magnetization profile to Eq.~\ref{Eq:prof_zeta}. 
Equation~\ref{Eq:zeta} and the discrete Heisenberg model give very similar results for $\zeta$.

For large perpendicular magnetic fields, $\zeta$ approaches a maximum of $\pi/4$. 
This value corresponds to full magnetization saturation along the direction of applied magnetic field. 
As a result of diminishing magnetization rotation between domains ($\Delta$), 
the DW width ($\lambda'$) decreases with increasing field strength (Fig.~\ref{Fig:dw_happ}(b)). 
The predicted tunability of the width and internal spin structure 
of a pinned DW might be exploited for active manipulation of spin waves. 
Previously, it has been found that dynamic stray fields in DWs reduce the transmission of 
propagating spin waves if the DW width becomes smaller than the spin-wave wavelength.~\cite{MAC-10,WAN-APL-13} 
Reprogramming of the DW spin structure by an external field at a fixed location of a ferromagnetic film could thus 
impose controllable changes to the amplitude or phase of passing spin waves, 
which is an essential feature of magnonic logic devices.~\cite{CHU-15}  

\section{Current-induced domain wall dynamics}
\label{Sec:Dynam}

\subsection{Zero magnetic field}

\begin{figure}[!htp]
 \centering
 \includegraphics[width=.9\columnwidth]{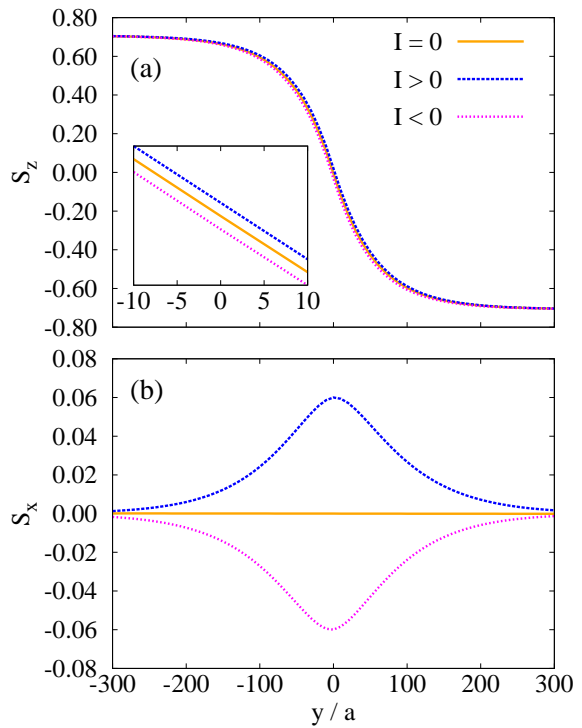}
 \caption{Head-to-tail DW profiles under the influence of an electric current for (a) in-plane, $S_z$, 
 and (b) out-of-plane, $S_x$, spin coordinates as obtained from Heisenberg model simulations. 
 The inset of (a) shows the displacement of the DW center from the anisotropy boundary at $y=0$. 
 In the calculations $\alpha = 0.15$, $P=0.5$, $\beta=0.4$, and $|I| = 10^{12}\,{\rm A}/{\rm m}^2$. 
 The other parameters are the same as in Fig.~\ref{Fig:DW_prof}.}
 \label{Fig:DW_curr}
\end{figure}
We now discuss the influence of an electric current on the DW profile and its dynamic properties. 
We first focus on a pinned head-to-tail 90$^\circ$ DW in zero magnetic field. 
To examine the action of spin-transfer torque, we perform numerical simulations of the Heisenberg model with constant spin current density. 
In these simulations, we develop the magnetization using the LLG equation until a stationary state is reached. 
Results for a current density of $I = \pm 10^{12}\, {\rm A}/{\rm m}^2$, 
which is comparable to values used in experiments,~\cite{YAM-04,Klaui2005:PRL} are shown in Fig.~\ref{Fig:DW_curr}. 
While the electric current does not substantially modify the in-plane DW profile, 
it shifts the DW center away from the magnetic anisotropy boundary (see inset in Fig.~\ref{Fig:DW_curr}(a)). 
Moreover, the DW magnetization tilts out of the film plane under the action of an electric current (Fig.~\ref{Fig:DW_curr}(b)). 
The direction of DW displacement and sign of DW tilt depend on the direction of electric current. 
The magnitude of both effects are determined by the absolute value $|I|$. 
Our results are consistent with current-induced magnetization dynamics of 180$^{\circ}$ DWs.~\cite{Li:PRL_2004,Li:PRB_2004} 
Importantly, the results of Fig.~\ref{Fig:DW_curr} allow us to assume that the profile of a pinned 90$^{\circ}$ DW 
does not change under the influence of an electric current. 
Hence, we can describe DW dynamics by two collective coordinates, namely, 
the position of the DW center ($q(t)$) and the DW tilt angle ($\psi(t)$), 
as illustrated in Fig.~\ref{Fig:collective}.

In agreement with Eq.~\ref{Eq:prof_eq}, we use the following ansatz for the in-plane DW profile
\begin{equation}
  \tilde{\theta}(y,t) =
    \begin{dcases}
      \frac{\pi}{4} + 2\, \arctan\left[ \left( \sqrt{2} - 1 \right) \exp\left(\frac{y-q(t)}{\lambda} \right) \right]\,, \\
      \qquad \text{if } y < q(t)\,, \\
      \frac{3 \pi}{4} - 2\, \arctan\left[ \left( \sqrt{2} - 1 \right) \exp\left(-\frac{y-q(t)}{\lambda} \right) \right]\,. \\
      \qquad \text{if } y > q(t)\,.
    \end{dcases}
\label{Eq:theta_ansatz}
\end{equation}
The out-of-plane DW profile needs to satisfy vanishing magnetization and spin-transfer torque inside the domains, i.e., 
far away from the anisotropy boundary. To account for this, we use
\begin{equation}
  \tilde{\phi}(y,t) = \frac{\pi}{2} - \psi(t)\, \cos\left[2\, \tilde{\theta}(y,t) \right]\,.
\label{Eq:phi_ansatz}
\end{equation}
Here, the DW tilting angle $\psi$ corresponds to the maximum out-of-plane magnetization angle. 
We note that this simple ansatz does not fully reproduce the numerical simulations of Fig.~\ref{Fig:DW_curr}(b). 
In Eq.~\ref{Eq:phi_ansatz}, $\phi$ decays more quickly as a function of $y$ compared to the Heisenberg model. 
Despite this discrepancy, we will demonstrate that the approximation is valid 
for calculations of the tilting angle and resonance frequency in the limit of small DW displacements.

We obtain dynamic equations for the collective DW coordinates, $q(t)$ and $\psi(t)$, 
by using $\delta{w}/\delta{\theta}$ and $\delta{w}/\delta{\phi}$ 
from Eq.~\ref{Eqs:LLG_spher} and defining the differential areal energy density
\begin{equation}
  {\rm d}\varepsilon = 
  \int_{-\infty}^{\infty} {\rm d}y \left[ \left(\fder{w}{\theta}\right)\delta\theta + \left(\fder{w}{\phi}\right)\delta\phi \right]\,.
\label{Eq:deps}
\end{equation}
Inserting Eq.~\ref{Eq:theta_ansatz} and Eq.~\ref{Eq:phi_ansatz} into Eq.~\ref{Eq:deps} and 
integrating along $y$ gives an equation of motion for the collective coordinates
\begin{equation}
  \der{}{}{t} \begin{pmatrix} q \\ \psi \end{pmatrix} =
  \bar{\bm M}\,
  \begin{pmatrix}
    \partial \epsilon / \partial q\\
    \partial \epsilon / \partial \psi
  \end{pmatrix} +
  \begin{pmatrix}
    a_u \\
    b_u
  \end{pmatrix}\, u\,,
\label{Eq:motion}
\end{equation}
where
\begin{equation}
  \renewcommand*{\arraystretch}{1.9}
  \bar{\bm M} = -\frac{\gamma}{\Ms}\, \frac{3}{2 \sqrt{2}}
                \begin{pmatrix}
		  \frac{5\sqrt{2}+1}{5}\, \alpha\, \lambda & -1 \\
                  1 & \frac{3}{2} \frac{\sqrt{2} - 1}{\lambda}\, \alpha
                \end{pmatrix}\,,
\label{Eq:matM}
\end{equation}
and
\begin{subequations}
  \begin{align}
    a_u &= 1 + \alpha \beta\,, \\
    b_u &= -\frac{3 (\sqrt{2} - 1)}{2}\, \frac{\alpha - \beta}{\lambda}\,.
  \end{align}
\label{Eqs:curr_amps}
\end{subequations}

Equation~\ref{Eq:motion} can be linearized. This gives
\begin{equation}
  \der{}{}{t} \begin{pmatrix} q \\ \psi \end{pmatrix} =
  \bar{\bm D}\,
  \begin{pmatrix}
    q\\
    \psi
  \end{pmatrix} +
  \begin{pmatrix}
    a_u \\
    b_u
  \end{pmatrix}\, u\,,
\label{Eq:motion_lin}
\end{equation}
where $\bar{\bm D}$ is the dynamic matrix
\begin{equation}
  \bar{\bm D} = \bar{\bm M} \cdot 
  \begin{pmatrix}
    \partial^2 \epsilon / \partial q^2 & \partial^2 \epsilon / \partial q \partial \psi \\
    \partial^2 \epsilon / \partial \psi \partial q & \partial^2 \epsilon / \partial \psi^2
  \end{pmatrix}_{\!\rm eq}\,,
\label{Eq:dynam_approx}
\end{equation}
where the subscript ${\rm eq}$ indicates that the second derivatives of the areal energy density 
are evaluated numerically in the equilibrium magnetic configuration, 
i.e., $q = 0$ and $\psi = \pi/2$.~\cite{Voto2017:SciRep,Bazaliy2004:PRB}

\begin{figure}[!tbp]
 \centering
 \includegraphics[width=.9\columnwidth]{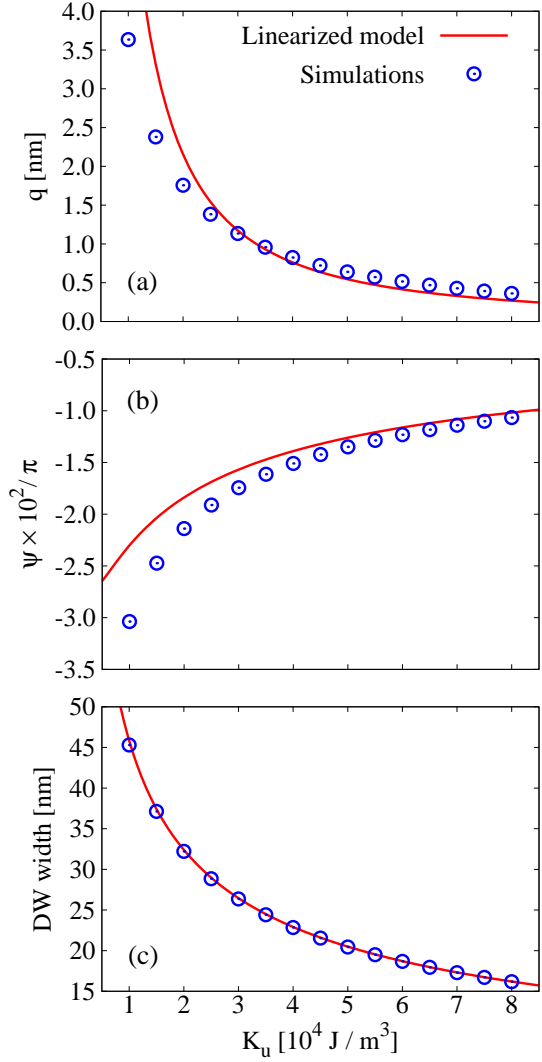}
 \caption{Comparison of the linearized analytical model and Heisenberg model simulations for 
          $\alpha = 0.15$, $P=0.5$, $\beta=0.4$, and $I = 10^{12}\, {\rm A}/{\rm m}^2$. 
          The other parameters are the same as in Fig.~\ref{Fig:DW_prof}.
          (a) DW displacement, (b) DW tilting angle, (c) DW width.
          In (c), the DW width obtained from Heisenberg model simulations is compared to Eq.~\ref{Eq:lambda}.}
 \label{Fig:model_compar}
\end{figure}

Let us now discuss the validity and applicability of the linearized 1D model. 
Figure~\ref{Fig:model_compar} compares the stationary values of $q$ and $\psi$ 
under constant electric current as a function of $K_u$. As $K_u$ increases, 
both the DW displacement and DW tilting angle decrease because of stronger pinning at the anisotropy boundary. 
As a result, the linearized 1D model is more accurate for large values of $K_u$. 
This is confirmed by Figs.~\ref{Fig:model_compar}(a) and (b), 
where the parameter values of the 1D model approach the numerical simulations when the anisotropy is strong.
In addition, if we fit the displaced DW profile with Eq.~\ref{Eq:theta_ansatz}, 
we obtain a DW width that is comparable to Eq.~\ref{Eq:lambda} in the whole anisotropy range (Fig.~\ref{Fig:model_compar}(c)). 
Based on these results, we conclude that our linearized 1D model describes current-induced DW dynamics 
in the approximation of small DW displacements. 
The calculated DW displacement for a current density $I = 10^{12}\, {\rm A/m^{2}}$ is of the order $\sim$1 nm. 
This distance compares well to micromagnetic simulations in Ref.~\citenum{VanDeWiele:SciRep_2016}. 
In the same study it was shown that DW oscillations of this amplitude, driven by an ac spin-polarized current, 
turn the pinned 90$^{\circ}$ DW into a tunable source of propagating spin waves.

Now, we discuss the effect of an ac electric current in our model. 
Since the direction of DW displacement depends on the direction of current, 
an ac electric current induces DW oscillations around its equilibrium position. 
For potential applications in magnonics, the DW resonance frequency ($\omega_{\rm r}$) is a key parameter.~\cite{Saitoh2004:Nat}
To calculate $\omega_{\rm r}$, we use the linearized equations of motion (Eq.~\ref{Eq:motion_lin}). 
For an ac electric current with frequency $\omega$, we write $I(t) = I_0\, e^{-i\, \omega t}$. 
Moreover, we assume that the solutions of the linearized equation of motion have the same form, 
$q(t) = C_q\, e^{-i\, \omega t}$ and $\psi(t) = C_\psi\, e^{-i\, \omega t}$, where $C_q$ and $C_\psi$ are constants. 
Using these parameters, we find that Eq.~\ref{Eq:motion_lin} has a solution for a DW resonance frequency of
\begin{equation}
  \omega_{\rm r} = \frac{3}{2 \sqrt{2}}\, \frac{\gamma}{\Ms}  
  \sqrt{\pderr{2}{\varepsilon}{q}\pderr{2}{\varepsilon}{\psi} - \left( \ppder{\varepsilon}{q}{\psi} \right)^2}\,,
\label{Eq:om_res}
\end{equation}
where, as in the Eq.~\ref{Eq:dynam_approx}, the second derivatives of the areal energy density ($\varepsilon$) 
are evaluated numerically in the equilibrium magnetic configuration. 

The solid line in Fig.~\ref{Fig:res_freq}(a) shows the dependence of 
$f_{\rm res} = \omega_{\rm r} / (2\pi)$ on $K_u$ in the absence of a magnetic field. 
We find that $\omega_{\rm r} \sim K_u^{1/2}$. 
In addition, the potential stiffness, $\kappa$, which is defined by $\partial\varepsilon/\partial{q} = \kappa\, q$ 
can be approximated as $\kappa = K_u/\lambda$. 
This gives $\kappa \sim K_u^{3/2}$. 
Finally, the DW mass $m_{\rm DW} = \kappa / \omega_{\rm r}^2$,~\cite{Dorig1948:ZNat,Saitoh2004:Nat} 
which can be used as an indicator for the operation speed of DW devices, varies as $m_{\rm DW} \sim K_u^{1/2}$.

\subsection{Simultaneous effect of magnetic field and electric current}

In Sec.~\ref{SSec:mag_field} we showed that an in-plane magnetic field along the $y$-axis 
reduces the magnetization rotation between domains ($\Delta$) and the DW width ($\lambda'$). 
This might also modify the DW resonance frequency. 
By combining the expression for zero-field resonance frequency (Eq.~\ref{Eq:om_res}) 
and ansatzes for the DW profile (Eqs.~\ref{Eq:prof_zeta} and~\ref{Eq:phi_ansatz}), 
we derive dynamic equations for the collective coordinates in non-zero magnetic fields. 
The equation of motion has the same form as Eq.~\ref{Eq:motion} with $\bar{\bm M}$ replaced by 
\begin{equation}
  \renewcommand*{\arraystretch}{1.9}
  \bar{\bm M}(\zeta) = \frac{\gamma}{\Ms}
                \begin{pmatrix}
		 \alpha \lambda'\, f(\zeta)^{-1} & g(\zeta)^{-1} \\
                  -g(\zeta)^{-1} & \alpha\, h(\zeta)^{-1} / \lambda'
                \end{pmatrix}\,,
\label{Eq:matM}
\end{equation}
and
\begin{subequations}
  \begin{align}
    a_u &= 1 + \alpha \beta\,, \\
    b_u &= \frac{\alpha - \beta}{\lambda'}\, \frac{f(\zeta)}{g(\zeta)}\,.
  \end{align}
\label{Eqs:curr_amps}
\end{subequations}
Here, the three functions that vary with $\zeta$ are given by
\begin{subequations}
  \begin{align}
    f(\zeta) = &\sqrt{2}\, \left( \sin\zeta + \cos\zeta - \sqrt{2} \right)\,, \\
    g(\zeta) = &\frac{2 \sqrt{2}}{3}\, \frac{\sin\zeta + \cos\zeta - \sqrt{2}\sin(2\zeta)}{\cos(2\zeta)}\,, \\
    h(\zeta) = &\frac{\sqrt{2}}{15} \frac{1}{\cos^2(2\zeta)}\, \biggl[ 5\, (\sin\zeta + \cos\zeta)\; - \\
               & 2\sqrt{2}\, \sin(2\zeta) + 7\, [\sin(3\zeta) - \cos(3\zeta)] - 10\sqrt{2}\; \biggr]\,. \notag
  \end{align}
\label{Eqs:abc}
\end{subequations}
After linearization, we obtain an expression for the DW resonance frequency as a function of magnetic field
\begin{equation}
  \omega_{\rm r}(\zeta) = 
  \frac{\gamma}{\Ms}\, g(\zeta)^{-1}\,  
  \sqrt{\pderr{2}{\varepsilon}{q}\pderr{2}{\varepsilon}{\psi} - \left( \ppder{\varepsilon}{q}{\psi} \right)^2}\,.
\label{Eq:om_res_happ}
\end{equation}
Here, we applied the approximate relation $f(\zeta) h(\zeta) \simeq g^{2}(\zeta)$. 
The DW resonance frequency depends on the function $g(\zeta)^{-1}$. 
For zero applied field $g(0)^{-1} = 3 / (2\sqrt{2})$, which recovers Eq.~\ref{Eq:om_res}. 
$g(\zeta)^{-1}$ increases with $\zeta$ and diverges for $\zeta \to \pi/4$, i.e., when the DW is erased by the applied magnetic field.

Figure~\ref{Fig:res_freq}(b) shows the field-dependence of 
$f_{\rm res}(\zeta) = \omega_{\rm r}(\zeta)/(2\pi)$ for several values of $K_u$. 
The resonance frequency increases as a function of $\happ$. 
This effect relates to a reduction of the DW width at nonzero $\happ$ (see Fig.~\ref{Fig:dw_happ}(b)). 
For narrow DWs, the stiffness of the pinning potential increases, causing an upshift of $f_{\rm res}$. 
Our calculations indicate nearly linear tuning of $f_{\rm res}$ by several GHz in modest magnetic fields. 
This ability to actively alter $f_{\rm res}$ could be used to tailor the frequency and wavelength 
of spin waves that are emitted from an oscillating DW.

\subsection{Comparison with micromagnetic simulations}

\begin{figure}[!tbp]
 \centering
 \includegraphics[width=.9\columnwidth]{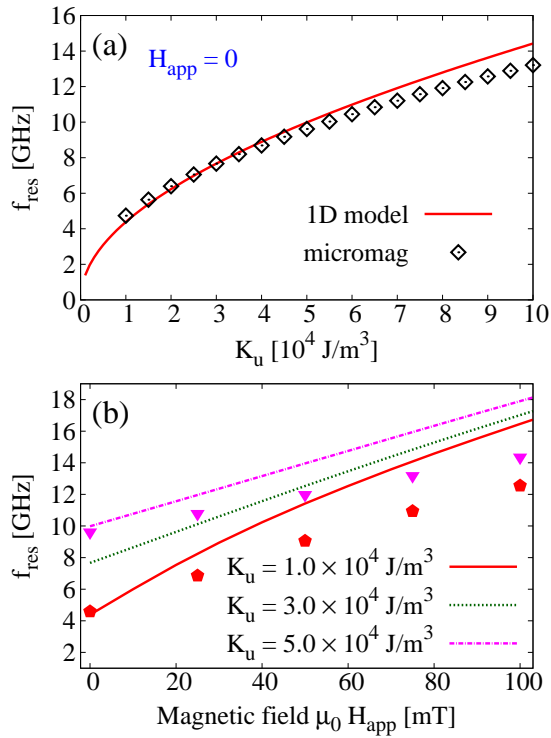}
 \caption{(a) DW resonance frequency calculated for $\happ = 0$. 
          The line is calculated using the 1D analytical model and the open diamonds are obtained from micromagnetic simulations. 
          (b) Field dependence of the DW resonance frequency calculated using the 1D model (lines) for various values of 
          $K_u$ and extracted from micromagnetic simulations (solid symbols) using
          $K_u = 1.0\times 10^4\, {\rm J}/{\rm m}^3$ (pentagons)  and $5.0\times 10^4\, {\rm J}/{\rm m}^3$ (triangles).}
 \label{Fig:res_freq}
\end{figure}
In the previous sections, we derived a 1D analytical model for a magnetic DW that is pinned by a 90$^\circ$ uniaxial anisotropy boundary. 
Results from this model for the DW profile, DW displacement, 
and DW tilting angle were compared to numerical simulations based on the 1D Heisenberg model. 
Although the 1D Heisenberg model goes beyond a simple linear approximation and the assumption of a rigid DW profile, 
it might deviate from reality because of its reduced dimensionality and 
lack of long-range dipolar interactions.~\cite{Beach2008:JMMM,Vandermeulen2018:JMMM}
Therefore, we will now compare our model results 
to micromagnetic simulations and assess its relevance for the interpretation of experimental data.

The simulations were performed using MuMax3 software~\cite{mumax3} with periodic boundary conditions in the y-z plane. 
Modulations of uniaxial magnetic anisotropy were included by abrupt rotation of the magnetic easy axis 
at the cell boundary of two 10-$\mu$m-wide stripe domains. 
The film thickness was set to 5 nm and the structure was discretized into $2.44 \times 4.88 \times 5\, {\rm nm}^3$ cells. 
We estimated the resonance frequency of the pinned DW by applying 
a ${\rm sinc}$-function-type current pulse in the $y$-direction with a cut-off frequency of 
$40\, {\rm GHz}$. After this, the $z$-component of magnetization was recorded one cell from the anisotropy boundary. 
The eigen frequency of the DW was extracted by performing a Fourier transformation on these data. 

\begin{figure}[htp!]
  \centering
  \includegraphics[width=.9\columnwidth]{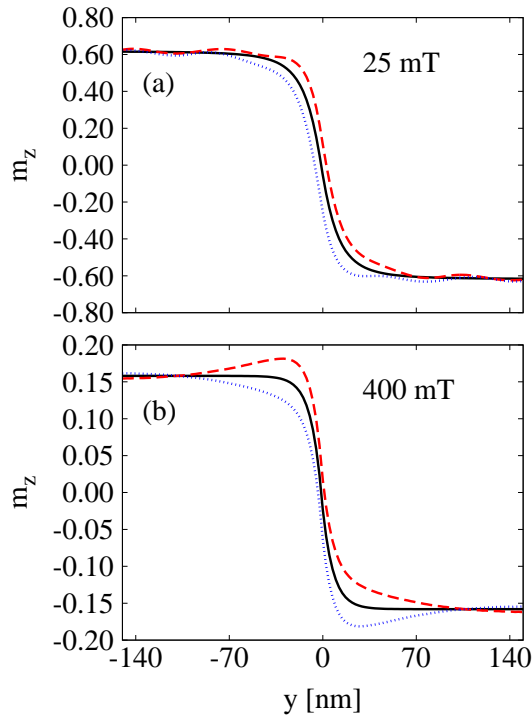}
  \caption{Micromagnetic simulations of the DW profile during current-induced magnetization dynamics. 
  Panels (a) and (b) show results for different magnetic bias fields along the $y$-axis. 
  The solid black lines depict DW profiles in equilibrium (zero current). 
  The dashed red and dotted blue lines show snapshots of the displaced and distorted DW during current-driven oscillations. 
  The anisotropy constant in the simulations is $K_u = 10^5\, {\rm J}/{\rm m}^3$.}
\label{Fig:dw_excit}
\end{figure}

The simulated profile and width of the pinned DW in zero and non-zero magnetic field agree well with results from our 1D model. 
The main effect of dipolar interactions, which are included in the micromagnetic simulations but omitted in the 1D model, 
is an enlargement of the DW tails. 
We also find good correspondence between the simulated and calculated values of the DW resonance frequency. 
Figure~\ref{Fig:res_freq}(a) shows a comparison for different values of $K_u$ and zero magnetic field. 
At large magnetic field, the results start to deviate, as shown in Figs.~\ref{Fig:res_freq}(b). 
Under these conditions, the 1D model overestimates the DW resonance frequency. 
One of the reasons is a gradual decrease of the magnetization rotation between domains ($\Delta$). 
This effect lowers the spin-transfer torque efficiency and thereby the displacement of the DW. 
Another factor relates to a distortion of the DW during magnetization dynamics. 
The dependence of both effects on applied magnetic field is illustrated in Fig.~\ref{Fig:dw_excit}. 
The figure shows micromagnetic simulations of the displacement and deformation of the DW during current-induced DW oscillations. 
The applied magnetic field in (a) and (b) is $\mu_0\happ = 25$~mT and $\mu_0\happ = 400$~mT, respectively. 
The solid black lines represent DW profiles for zero electric current and the other lines depict snapshots of dynamic DW deformations. 
In small magnetic field, the spin-transfer torque displaces the DW without significantly changing its profile. 
Because of smaller spin-transfer torque efficiency, 
the DW displacement diminishes upon an increase of the magnetic field strength. 
At the same time, deformations of the DW profile become more pronounced. 
Because our 1D analytical model assumes a rigid DW, it overestimates the DW resonance frequency for large magnetic field.  

\section{Conclusions}
\label{Sec:Conclusions}

In summary, we studied the static and dynamic properties of a magnetic DW that is pinned by a 90$^\circ$ uniaxial anisotropy boundary 
using an analytical model with continuous spatial coordinates and a discrete Heisenberg model. 
First, we derived a formula for the profile of an equilibrium head-to-tail DW. 
To account for the abrupt rotation of magnetic anisotropy, 
we split the expression for the in-plane magnetization profile into two parts (Eq.~\ref{Eq:prof_eq}). 
Consequently, calculations for the two domains were done separately. We note that the following ansatz can be used to simplify the model
\begin{equation}
  \theta_{{\rm u},{\rm approx}}(y) = \frac{\pi}{4} + \arctan\left[ \exp \left( \frac{y}{\lambda} \right) \right]\,.
\label{Eq:prof_approx}
\end{equation}
Here, $\lambda$ is given by Eq.~\ref{Eq:lambda}. Equation~\ref{Eq:prof_approx} does not satisfy Eq.~\ref{Eq:dif_eq}, 
but its similar shape could be sufficient for practical purposes. 
After assessing the equilibrium state, we analyzed how the DW profile deforms in a magnetic field. 
Besides an obvious reduction of the magnetization rotation between domains, 
we observed a gradual decrease of the DW width in a perpendicular magnetic field. 

Next, we used the Landau-Lifshitz-Gilbert equation to explore current-induced dynamics of a pinned DW. 
For a small electric current and zero magnetic field, 
we found that the DW is slightly displaced from the anisotropy boundary without significantly changing its in-plane magnetization profile. 
Additionally, the spin-transfer torque tilts the DW magnetization out of the film plane. Using an ansatz for the DW profile, 
we derived linear equations of motion for collective DW coordinates and demonstrated that the calculated values of 
DW displacement and DW tilting angle are in good agreement with Heisenberg model simulations. 
We also derived expressions for the DW resonance frequency in zero and non-zero magnetic fields. 
Our results indicate that an ac electric current can drive the domain wall into resonance. 
Moreover, the model predicts active tuning of the DW eigen frequency by a magnetic bias field. 
Finally, we showed that our model calculations are in good agreement with micromagnetic simulations up to modest magnetic fields. 
Beyond this, break-down of the rigid-DW approximation causes an overestimation of the DW resonance frequency.  

Spin waves are emitted from a pinned DW if an ac spin-polarized current or another activation mechanism forces it to oscillate. 
To exploit DW pinning at anisotropy boundary in programmable magnonic devices one needs to understand
their basic static and dynamic properties and learn how to control them.
The models provided here describe active tuning of the DW resonance condition by
means of an external magnetic field.

\section*{Acknowledgement}
This work was supported by the European Regional Development Fund 
in the IT4Innovations national supercomputing center - path to exascale project 
(project number CZ.02.1.01/0.0/0.0/16\_013/0001791 within the Operational Programme Research, Development and Education) 
and the European Research Council (grant number ERC-2012-StG 307502-E-CONTROL). 
PB thanks the Czech Science Foundation for support (grant number 18-07172S) 
and S.J.H. acknowledges support from the V\"ais\"al\"a Foundation. 
The micromagnetic simulations were performed using computational resources provided by the Aalto Science-IT project.



\end{document}